\documentclass[letter,scriptaddress,twocolumn, prd,showkeys]{revtex4}

	\usepackage{amsmath}
	\usepackage{makeidx}
	\usepackage{amsfonts}
	\usepackage[ansinew]{inputenc}
	\usepackage[usenames,dvipsnames]{pstricks}
	\usepackage{subfigure}
	\usepackage{epsfig}
	\usepackage{pst-grad} 
	\usepackage{pst-plot} 
	\usepackage[colorlinks,hyperindex]{hyperref}
	\hypersetup
	{
		colorlinks,%
		citecolor=black,%
		linkcolor=black,%
		urlcolor=black,%
	}

\makeindex

\begin{document}

\title{The non-integrable mass and the scalar charge}

\author{Qiang Wen}
\email{qwen@math.tsinghua.edu.cn}
\affiliation{Yau Mathematical Sciences Center, Tsingshua University, Beijing 100084, China}

\affiliation{Perimeter Institute for Theoretical Physics, Waterloo, Ontario N2L 2Y5, Canada}
\affiliation{Department of Physics, Renmin University of China, Beijing 100872, China.}

\date{\today}

\begin{abstract}
The non-integrable mass is studied explicitly in this paper. We study Einstein-scalar gravities with weakened boundary conditions, and calculate the mass with the Hamiltonian formula and Wald's formula respectively. We find the masses calculated by these two formulas are non-integrable. One way to solve this non-integrability problem is to impose boundary conditions; however, we find the mass calculated in this way has many other problems. This implies the macroscopic thermodynamic properties of the scalar hairy black holes should be described by one more charge beside the mass, which we call a scalar charge. In fact, the non-integrability of mass will always arise when the matter fields have charges which is not associate to any diffeomorphisms of spacetime. We find the mass becomes non-integrable just because Wald's formula is used in a wrong way.  Based on Wald's formula and the existence of the scalar charge, we propose a new definition for mass, with the modification that we require the variation of the Hamiltonian to have no contribution from the variation of the other charges. This new definition is also valid for much more general gravities coupled to matter fields with other charges.
\end{abstract}

\maketitle

\section{Introduction}
Although asymptotically AdS spacetimes (in this paper, when we refer asymptotically AdS spacetimes we mean spacetimes asymptotically goes to AdS, whose boundary conditions donot have to preserve all the asymptotic AdS symmetries)  have attracted remarkable interest in recent decades, our understanding of their conserved charges, like energy, is still not clear enough, especially when gravity is coupled to matter fields and the boundary conditions cannot preserve all the
asymptotic AdS symmetries. Several method have been developed to define conserved charges in asymptotically $AdS_{d}$ spacetimes: the Hamiltonian definition developed by Henneaux and Teitelboim~\cite{Henneaux:1985tv}, the AMD's (Ashtekar-Magnon-Das) conformal method~\cite{Ashtekar:1984zz,Ashtekar:1999jx}, the ``counterterm subtraction method"~\cite{Henningson:1998gx,Balasubramanian:1999re,Skenderis:1999nb,Bianchi:2001de,deHaro:2000wj,Skenderis:2002wp,de Haro:2000xn,Skenderis:2000in,Bianchi:2001kw,Papadimitriou:2004ap}. There are also other methods like the KBL method ~\cite{Katz:1996nr,Deruelle:2004mv}, the spinor method~\cite{Deruelle:2004mv,Gibbons:1982jg} and the ``pseudotensor" method~\cite{Abbott:1981ff}, which we will not consider further in this paper.  Based on the general ``covariant phase space formalism" developed by Wald et. al.~\cite{Wald:1993nt,Iyer:1994ys}, Hollands, Ishibashi and Marolf developed another method~\cite{Hollands:2005wt} to calculate the conserved charges in asymptotically AdS-spacetimes, and made a comparison between those methods mentioned above. They showed that, with boundary conditions preserving all the asymptotic AdS symmetries and the matter fields decreasing fast enough when approaching the boundary, those methods are all equivalent to each other, except the energy calculated by ``counterterm subtraction method" has an additional constant term in some cases, which can be interpreted as the Casmir energy of the dual CFT.

These methods all have strict requirements: not only the boundary conditions should preserve all the asymptotic AdS symmetries, but also the stress-tensor or matter fields should decrease fast enough when approaching boundary. However, in general these requirements are not satisfied by the asymptotic behaviors of the solutions, and thus, the boundary conditions are weakened. For example, consider gravity minimally coupled to a scalar field $\phi$ with a scalar mass $m$ above or saturating the BF (Breitenlohner-Freedman) bound~\cite{BF}, with the Lagrangian (note that in this paper we use the convention $16\pi G=1$) expressed as
\begin{equation}\label{Lag}
\mathcal{L}=\sqrt{-g}\left(R-\frac{1}{2}(\partial \phi)^2-V(\phi)\right).
\end{equation}
It is known that, the tachyonic ($m^2<0$) scalars AdS spacetime are stable provided their mass is above the BF bound. In these theories the asymptotic AdS symmetries cannot be fully preserved by the boundary conditions in general (which we will see later), and also the scalar field usually decreases slower than the requirements mentioned above. One can find many examples in~\cite{Lu:2014maa}, where the asymptotic behaviors of many different solutions are studied. 

If we consider a scalar mass slightly above the BF bound and use the static spherical metric ansatz,
\begin{equation}\label{M1}
ds^2=-h(r)dt^2+\frac{dr^2}{f(r)}+r^2 d\Omega_{d-2}^{2}\,,\qquad \phi=\phi(r)\,,
\end{equation}
where $d$ is the spacetime dimension and $d\Omega_{d-2}^{2}$ is the metric on the unit $(d-2)$-sphere, then the asymptotic behaviour of the scalar field is given by
\begin{align}
  \phi(r)=\frac{\phi_1}{r^{\lambda_+}}+\frac{\phi_2}{r^{\lambda_-}}+\cdots\,.
 \end{align}
The two parameter which satisfy $\lambda_+>\lambda_-$, are determined by the spacetime dimentsion $d$ and the scalar mass $m^2$ (see (\ref{sigma}) and (\ref{sle})). When $\phi_1\neq 0$, the speed requirement for the decreasing of the scalar field cannot be satisfied. And the asymptotic AdS symmetries cannot be fully preserved unless we impose a family of special boundary conditions between $\phi_1$ and $\phi_2$~\cite{Henneaux:2004zi,Hertog:2004dr,Henneaux:2006hk}, which are:
\begin{align}\label{boundarycondition}
a)~~\phi_1&=0\,,
\cr
b)~~\phi_2&=0\,,
\cr
c)~~\phi_2&=\mathcal{C}\,\phi_1^{\lambda_-/\lambda_+}\,,
\end{align}
where $\mathcal{C}$ is an arbitrary constant with out variation.

With their requirements not satisfied, the methods mentioned above begin to give different masses. Although they still give finite results in many cases (for details one should consult~\cite{Lu:2014maa}). We can not trust their results in these situations. 

Under such weakened boundary conditions, some work~\cite{Henneaux:2004zi,Hertog:2004dr,Henneaux:2006hk} was done to calculate the conserved charges in Einstein-scalar gravities using the the Hamiltonian formula (or the Henneaux-Teitelboim construction). Soon after, the same results ~\cite{Hertog:2005hm,Amsel:2006uf} were obtained using Wald's formula~\cite{Wald:1993nt,Iyer:1994ys}.  The only requirement on boundary conditions for the Hamiltonian formula and Wald's formula is to generate finite conserved charges, so it seems their results can be trusted. Their energy has contributions from both the gravity sector and the scalar field sector. Since the scalar field doesnot decrease fast enough when approaching the boundary, both of the two contributions have divergent terms. However, when we add them together, the divergent terms cancel with each other, thus, a finite result comes out in the end. In the next section we give a brief review on the Hamiltonian formula and Wald's formula, and show they are equivalent to each other in Einstein-scalar gravities.

However, the masses calculated in these two ways are non-integrable, or in other words the variation of the mass cannot be written as a total variation. This makes the energy ill-defined. We will explain why we call it ill-defined in section \ref{Pro}. In fact the non-integrability of mass also happens in gravities couples to other matter fields. In some cases, this problem can be solved by some proper gauge choice, for example the RN-AdS (Reissner-Nordstrom) black holes, however this does not work for the Einstein-scalar gravities. We also discuss another problem of these definitions of mass, which is related to the interpretation of the entropy of some hairy solutions. The validity of Wald's formula to calculate the conserved energy of scalar hairy solutions was discussed in~\cite{Hertog:2005hm,Amsel:2006uf}; however, we do not agree with their conclusions here. We use the RN-AdS black holes as an example to show that Wald's formula is in fact not used properly for calculating the energy of those scalar hairy solutions. 

Wald's formula is a very powerful method to study thermodynamics in gravities, especially when the gravities have coupled matter fields and weakened boundary conditions. Based on Wald's formula and with a little modification, we give a new definition of mass for these theories including Einstein-scalar gravities. Our definition is different from~\cite{Hertog:2005hm,Amsel:2006uf}. In order to definite the mass in our way, we need to introduce a new charge attached to the scalar field (section \ref{Sch}), we call this a scalar charge. The existence of this charge has already been discussed before from the black hole thermodynamics point of view~\cite{Gibbons:1996af,Lu:2013ura,Lu:2014maa,Liu:2013gja}. This definition gives the same results with the methods mentioned in the first paragraph when all their requirements are satisfied. And when the boundary conditions are weakened, the our results are different from the definitions given by~\cite{Henneaux:2004zi,Hertog:2004dr,Henneaux:2006hk,Hertog:2005hm,Amsel:2006uf} and do not have the problems mentioned in the previous paragraph.

We generalize our definition of mass to general gravities couple to matter fields in the Sec.
\ref{generalize}. And in the last section we further discuss the mass in gravity with boundary conditions.

\section{Hamiltonian formula and Wald's formula in Einstein-scalar gravities}
\subsection{The Hamiltonian formula}
The Hamiltonian formula directly construct the Hamiltonian of the theory~\cite{Regge:1974zd}, which is the usual Hamiltonian supplemented by an addition of a surface integral on the boundary
\begin{equation}
\textbf{H}=\int_{\Sigma}\xi^{\mu}\mathcal{H}_{\mu}+\text{surface integral terms}
\end{equation}
where $\Sigma$ is a spacelike surface and $\mathcal{H}_{\mu}$ are the constraints. Imposing the constraints
\begin{equation}
\mathcal{H}_{\mu}=0,
\end{equation}
the energy is just given by the surface integral terms. These surface integral terms are defined on a variational level, and have contributions from both the gravity sector, $\delta Q_{G}$, and the scalar field sector, $\delta Q _{\phi}$. Consider the Lagrangian given by (\ref{Lag}), we can calculate out these two contributions:
\begin{align}
&\delta Q_{G}=\int d{S_i}\bar{G}^{ijkl}(\xi^{\perp}\bar{D}_{j}\delta h_{kl}-\delta h_{kl}\bar{D}_{j}\xi^{\perp})\,,
\\
&\delta Q_{\phi}=-\int d{S_i} ~\xi^{\perp}\delta\phi D^{i}\phi\,,
\end{align}
where $G^{ijkl}=1/2 g^{1/2}(g^{ik} g^{jl}+g^{il}g^{jk}-2g^{ij}g^{kl})$, $h_{ij}=g_{ij}-\bar{g}_{ij}$ is the deviation from the pure AdS spatial metric $\bar{g}_{ij}$ and $\xi^{\perp}=\xi\cdot n$ with $n$ the unit normal to $\Sigma$.

Both $\delta Q_{G}$ and $\delta Q_{\phi}$ have divergent terms as the scalar field decays too slow when approaching the boundary; however, the divergent terms cancel with each other when adding them together~\cite{Henneaux:2004zi,Hertog:2004dr,Henneaux:2006hk}. Thus we get a finite variation of energy at last.

Consider the spherically metric $Ansatz$ (\ref{M1}), we find the variation of mass defined by this formula $\delta M_H=\delta Q_{G}+\delta Q_{\phi}$ is given by
\begin{equation}\label{Hf}
\delta M_H=-\omega_{d-2}r^2\sqrt{\frac{h}{f}}\left(\frac{d-2}{r}\delta f+f\phi'\delta \phi\right) \Big |_{r\rightarrow \infty}\,,
\end{equation}
where $\omega_{d-2}$ is the volume of the unit $(d-2)$-sphere. The $d=4$ case is given in~\cite{Hertog:2004bb}. One can get the mass by integrating out $\delta M_H$

\subsection{Wald's formula}
Here we write the covariant Lagrangian as a $d$-form $\textbf{L}$, and its variation can always be written in the form
\begin{equation}
\delta\textbf{L}=\textbf{E}\cdot\delta \varphi+d\Theta ,
\end{equation}
where $\Theta$ is a $(d-1)$-form, $\varphi$ represents all the metric functions and the scalar field. And $\textbf{E}$ represents the constraints, which will vanish when the equations of motion are satisfied. The Noether current defined by Wald and Iyer is defined as
\begin{equation}
\textbf{J}=\Theta-\xi\cdot \textbf{L}\,,
\end{equation}
which is shown~\cite{Lee:1990nz} to satisfy
\begin{equation}
d\textbf{J}=-\textbf{E}\mathcal{L}_{\xi}\phi\,.
\end{equation}
This means the current $\textbf{J}$ is conserved when $\varphi$ satisfies the equations of motion. Hence $\textbf{J}$ is a closed form and can be written as
\begin{equation}
\textbf{J}=dQ.
\end{equation}

It has been shown in~\cite{Wald:1993nt} that when $\xi$ is a Killing vector, we can get
\begin{equation}
d\delta Q-d(\xi\cdot\Theta)=0\,.
\end{equation}
This indicates the $(d-2)$-form $\delta Q-(\xi\cdot\Theta)$ is closed.

Stokes' theorem indicates that the integral of $(\delta Q_{\xi}-\xi\cdot\Theta)$ over a spherical surface is independent of the radius of the surface. In other words, the integral on the horizon $S_{r_0}$ and on the boundary $S_{\infty}$ are the same,
\begin{align}\label{Fl}
\int_{S_{r_0}}(\delta Q_{\xi}-\xi\cdot\Theta)=\int_{S_\infty}(\delta Q_{\xi}-\xi\cdot\Theta).
\end{align}
If one is familiar with Wald's formula, one would know that, the equation above usually gives the thermodynamic first law of the solutions. For pure gravity, when we take $\xi=\partial/\partial t$, the right hand side of (\ref{Fl}) gives $\delta M$ while the left hand side gives $T\delta S$, where $S=\frac{A}{4G}$ is proportional to the area $A$ of the event horizon. Thus, (\ref{Fl}) just gives the standard thermodynamic first law $T\delta S =\delta M$. 

The Wald's-formula definition of mass is just given by
\begin{equation}\label{Mw}
\delta M_W=\int_{S_\infty}(\delta Q_{\xi}-\xi\cdot\Theta)\Big|_{\xi=\partial/\partial t},
\end{equation}
where $M_W$ is considered as the conjugate charge associated to the Killing vector $\xi=\partial/\partial t$. The existence of this mass require the existence of a ($d-1$)-form $\textbf{B}$ which satisfies
\begin{equation}\label{Re1}
\delta \int_{S_\infty}\xi\cdot\textbf{B}=\int_{S_\infty} \xi\cdot\Theta\,,
\end{equation}
so the variation $\delta M_W$ can be written as a total variation. This makes sure that the mass can be integrated out with the result independent from the integral path. In~\cite{Wald:1999wa}, this requirement is reinterpreted as, for any variation $\delta_1$ and $\delta_2$ tangent to the space of solutions, the satisfaction of the following equation
\begin{equation}\label{Re2}
\int_{S_\infty} \xi\cdot \left(\delta_1\Theta(\varphi,\delta_2\varphi)-\delta_2\Theta(\varphi,\delta_1\varphi)\right)=0\,.
\end{equation}

Let us now consider the Einstein-scalar gravity (\ref{Lag}). We find the $(d-2)$-form $\xi\cdot\Theta$ also has contributions from both the gravity sector $\xi\cdot\Theta^{G}$, and the scalar field sector, $\xi\cdot\Theta^{\phi}$. After some calculations we get
\begin{align}
&\xi\cdot\Theta^{G}~_{i_1\cdots i_{d-2}}=\epsilon_{i_1\cdots i_{d-2} t \mu}(g^{\mu m}g^{\nu n}-g^{\mu n}g^{\nu m})D_{n}\delta g_{\nu m},
\\
&\xi\cdot\Theta^{\phi}~_{i_1\cdots i_{d-2}}=-\epsilon_{i_1\cdots i_{d-2} t }~^{\mu}(D_{\mu}\phi~\delta \phi),
\\
&Q_{i_1\cdots i_{d-2}}=\epsilon_{i_1\cdots i_{d-2}}~^{\mu t}D_{\mu}\xi_{t},
\end{align}
where $\epsilon$ is the Levi-Civita tensor. Applying the static metric ansatz (\ref{M1}) and choosing the Killing vector to be $\xi=\partial/\partial t$, we have~\cite{Lu:2014maa,Liu:2013gja}
\begin{align}\label{Wf}
\delta M_W
=-\omega_{d-2}r^{d-2}\sqrt{\frac{h}{f}}\left(\frac{d-2}{r}\delta f+f\phi'\delta\phi\right)\big |_{r\rightarrow \infty}\,,
\end{align}

Compare the two results (\ref{Wf}) and (\ref{Hf}), we see that, the Wald's-formula definition of mass is equivalent with the Hamiltonian-formula definition of mass in this case.

\section{Ill-defined mass}\label{Pro}
\subsection{Problem with a non-integrable mass}\label{ss}

Although the surface integrals on the boundary in the above two formulas are finite fro Einstein-scalar gravities, their definitions of mass have some fundamental problems. First the variation of their mass are actually non-integrable. 
 
We use the example in~\cite{Hertog:2004ns} to show this problem. Consider a 4-dimensional gravity (\ref{Lag}) with a scalar potential
\begin{align}\label{Sp}
V(\phi)=-2g^2(\cosh \phi+2),
\end{align}
where $g=1/\ell$. It can be shown that the scalar mass $m^2=-2$ is above the BF bound.
By solving the linearized equation of motion for the scalar field, the large radius decay of the scalar field is given by
\begin{equation}\label{Sa}
\phi(r)=\frac{\phi_1}{r}+\frac{\phi_2}{r^2}+\cdots.
\end{equation}

\begin{figure}

\centering
\includegraphics[width=0.5 \textwidth]{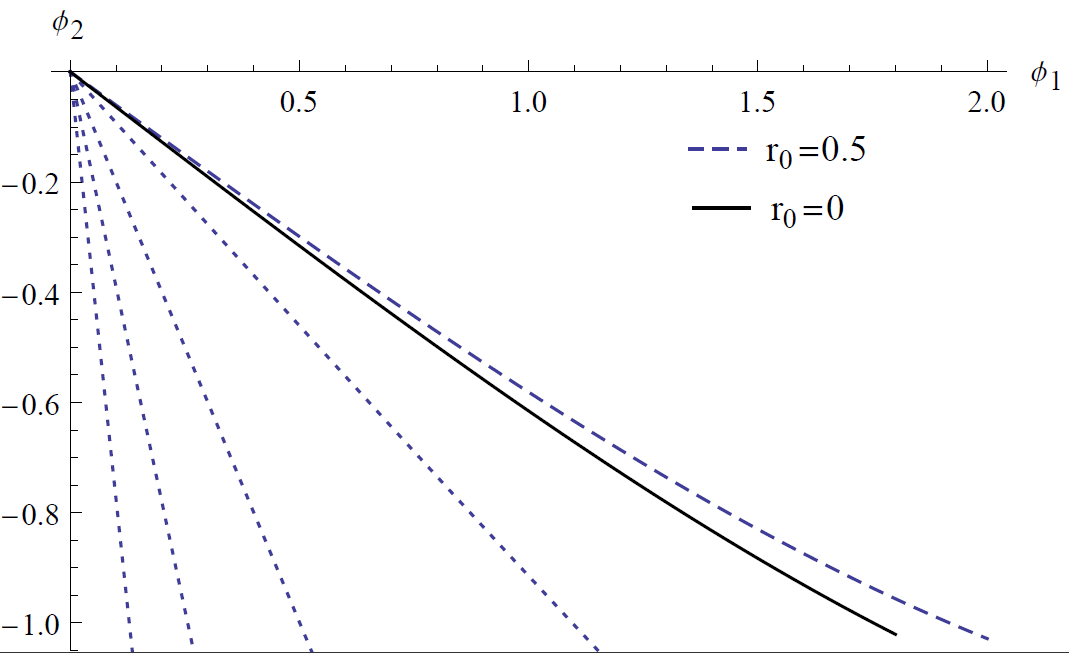}

\caption{\label{f1} The solid black line represents the line of soliton solutions with $r_0=0$. As $r_0$ increases the line of solutions goes right until to the dashed blue line which is the line of solutions with an event horizon around $r_0=0.5$. Continuing to increase $r_0$, the line of solutions then turns left, the four dotted lines from right to left representing lines of solutions with $r_0=2,5,10,20$. In the limit $r_0\rightarrow \infty$ the line of solutions approaches the $\phi_2$ axis. The orbit of the solution line gives the space of solutions on ($\phi_1$,$\phi_2$) plane}
\end{figure}

We now impose the metric ansatz (\ref{M1}), and numerically solve the equations of motion. By varying the initial values of the radius of the horizon $r_0$ and the value of the scalar field on the horizon $\phi(r_0)$, we get the space of solutions on the plane of ($\phi_1,\phi_2$), which are described by two independent parameters, see Fig. (\ref{f1}). Inside this space we can choose boundary conditions $\phi_2(\phi_1)$ freely. One thing we should keep in mind is that some points on ($\phi_1,\phi_2$) plane may represent many different solutions. In other words, it is not a one to one match between the plane and the space of solutions. For example the origin $(0,0)$ represents all Schwarzschild-AdS solutions with no scalar hair.

The energy $\delta M_H$ of these solutions is calculated in~\cite{Hertog:2004ns} with the Hamiltonian formula. Follow (\ref{Hf}), we get
\begin{equation}\label{Mh}
\delta M_{H}=4\pi\left(-2\delta \alpha-\delta (\frac{1}{3\ell^2}\phi_1 \phi_2)+\frac{\phi_2\delta\phi_1}{\ell^2}\right)\,,
\end{equation}
with $\alpha$ given by the large $r$ bahavior of $h(r)$,
\begin{equation}
h(r)=\frac{r^2}{\ell^2}+1+\frac{\alpha}{r}+\cdots.
\end{equation}

We see that $\delta M_H$ cannot be written as a total variation, hence is non-integrable. The directest way to solve this problem is assuming a boundary condition $\phi_2(\phi_1)$ thus giving an integral path. However this way will lead to controversail results. This means for a specific solution with specific metric and scalar field we cannot define its mass without imposing a boundary condition (or integral path) $\phi_2(\phi_1)$. Further more, the choice of boundary conditions in the space of solutions can be quite arbitrary and different boundary conditions will give different masses. This implies the mass of a specific solution is also arbitrary. This contradicts with the usual understanding of mass.

We can also use Wald's formula to calculate the mass. Applying to this case, the variation of mass calculated by Wald's formula (\ref{Mw}) agrees with (\ref{Mh}). The check the validity of the calculation, the holding of Eq. (\ref{Re2}) is also checked in~\cite{Hertog:2005hm,Amsel:2006uf}, provided we have assumed a boundary condition $\phi_2(\phi_1)$. Here we just cite the result, which is 
\begin{equation}\label{Re3}
\frac{\partial \phi_2(\phi_1)}{\partial\phi_1}(\delta_1\phi_1\delta_2 \phi_1-\delta_2\phi_1\delta_1\phi_1)=0.
\end{equation}

This means Eq. (\ref{Re2}) holds and it seems the calculation seems valid. However this does not mean the energy is independent of the integral path because when we choose a boundary condition $\phi_2(\phi_1)$, we have already chosen an integral path in the space of solutions. Hence we do not agree with that the holding of Eq. (\ref{Re3}) after choosing a boundary condition would make the definition of mass (\ref{Mw}) valid.

 \textit{We argue that, imposing a boundary condition to make a non-integrable mass integrable may be inappropriate}.

To further explain our arguments, we give a more convincing example, which is the calculation of the mass of four-dimensional electrically charged RN-AdS black holes (we set the magnetic charge to be zero for simplicity) using Wald's formula. The black hole metric is given by
\begin{align}\label{RN}
&ds^2=-f(r)dt^2+\frac{dr^2}{f(r)}+r^2 d\Omega_{2}^2\,,
\\
&f(r)=\frac{r^2}{\ell^2}+1-\frac{2 M}{r}+\frac{Q^2}{r^2}\,,
\end{align}
with two integration constants $M$ and $Q$, where $M$ is acknowledged as the black hole mass and $Q$ is the electric charge. We can make a gauge choice such that the gauge field $A$ vanishes on the horizon. Applying Wald's formula we find (for details, see~\cite{Gao:2003ys,Lu:2014maa})
\begin{align}
&\int_{S_{r_0}}(\delta Q_{\xi}-\xi\cdot\Theta)=T\delta S\,,
\\
\label{Mw2}
&\int_{S_{\infty}}(\delta Q_{\xi}-\xi\cdot\Theta)=\delta M-\Phi(M,Q)\delta Q=\delta M_W\,,
\end{align}
where $\Phi$ is the electrical potential. 
Wald's formula gives the standard thermodynamic first law $T\delta S=\delta M-\Phi \delta Q$. And like the case in Einstein-scalar gravities, $\delta M_W$ is also non-integrable here. If we assume a boundary condition $M(Q)$ again, we can integrate out $M_W$ and (\ref{Re2}) will hold. Following the logic of~\cite{Hertog:2005hm,Amsel:2006uf}, then the definition of mass (\ref{Mw}) seems valid here. However, in fact, the result is 
\begin{align}
M_W=\int \left(\frac{\partial M(Q)}{\partial Q}-\Phi(Q)\right)d Q
\end{align}
rather than the acknowledged mass $M$.

\subsection{The entropy problem of some massless solutions}
The Hamiltonian formula and Wald's formula definitions of mass have another problem, which is the failure to interpret the entropy of some scalar hairy solutions.

There are two ways to define the entropy of a black hole. The first is the definition from the thermodynamic point of view. The variation of the charges of the black hole satisfies a relationship (here is the Wald's formula), and if we consider the black hole as a thermodynamic system and interpret this relationship as the thermodynamic first law of the black hole, then there is a term which can play the role of entropy in thermodynamics. And the entropy is given by a simple formula $S=\frac{A}{4 G}$. This entropy is known as the Bekenstein-Hawking entropy.

The second definition is from the statistical point of view. It is widely believed that, a black hole is actually a quantum system with discrete energy levels~\cite{Susskind:1993if}, the total number of states which have the same macroscopic thermodynamic quantities is the exponential of the entropy. This definition is believed to be the microscopic origin of the first one, thus, both of them should give the same results. Inspired by the principle of holography, the quantum structures of a few black holes, for example the asymptotically $AdS_3$ black holes~\cite{Brown:1986nw,Strominger:1997eq} and extremal Kerr black holes~\cite{Guica:2008mu}, have already been studied, and the statistical entropy counted by Cardy's formula~\cite{Cardy:1986ie} do coincide with the Bekenstein-Hawking entropy. However, for more general black holes, it is still extremely hard to study their quantum structures.

In static Einstein-scalar gravities, if we define the mass with the Hamiltonian formula and Wald's formula, the mass $M$ is the only charge in the first law. Then the statistical entropy $S(M)$ is a function of only $M$, and counts the number of microscopic states which have a the same mass $M$.  However, we find some scalar hairy solutions whose Bekenstein-Hawking entropy and statistical entropy are not consistent with each other. This indicates that the definition of mass may be wrong.

The first example is the four-dimensional static spherical soliton solutions in~\cite{Hertog:2004ns} (and in Sec. \ref{ss}). We set $r_0=0$ and give different initial values for $\phi(0)$, then numerically solve the equations of motion. We get a series of soliton solutions (not black holes with $r_0=0$) which can be described by the soliton line (the solid black line) in Fig.(\ref{f1}). If we choose this soliton line as the boundary condition and define $\delta_s$ as the variation along this line, we should have $T\delta_s S=0$, since all the solutions on this line have no horizon. Thus Wald's formula (\ref{Fl}) gives
\begin{equation}\label{Dm}
\delta_s M_W=4\pi\left(-2\delta_s \alpha-\delta_s (\frac{1}{3\ell^2}\phi_1 \phi_2)+\frac{\phi_2\delta_s\phi_1}{\ell^2}\right)
=0\,.
\end{equation}
This indicates the masses $M_W$ of all the solitons are the same.
As the soliton line goes through the (0,0) point on ($\phi_1,\phi_2$) plane, which is the pure vacuum $AdS_4$ spacetime when there is no horizon, all the solitons are massless (if we define the $AdS_4$ vacuum as massless).

On one hand, the massless solitons have no horizon, hence no Bekenstein-Hawking entropy. While on the other hand, the number of states (or soliton solutions) with zero mass is absolutely more than one, thus their statistical entropy $S(M=0)$ should be nonzero. The inconsistence of the two entropies implies that the definition of mass given by Wald's formula might be wrong.

Another example is given by some 3-dimensional scalar hairy solutions constructed in~\cite{Wen:2015xea}. Theories (\ref{Lag}) with scalar potentials
\begin{subequations}\label{V1}
\begin{align}
&V(\phi)=g^2 \left(\cosh^{4\mu}\psi\right)\left(\mu\tanh^2\psi-2\right),
\\
&\psi=\frac{\phi}{2\sqrt{2\mu}},
\end{align}
\end{subequations}
admit massless (according to the Hamiltonian formula and Wald's formula) black hole solutions
\begin{subequations}\label{sl1}
\begin{align}
&\phi(r)=2\sqrt{2\mu}~\text{Arctanh}\frac{1}{\sqrt{1+r/q}}\,,
\\
&ds^2=-g^2 r^2 dt^2+\frac{1}{g^2 r^2(q/r+1)^{2\mu}} dr^2+r^2 d\theta^2\,,
\end{align}
\end{subequations}
with $\mu$ a positive constant which marks different theories and $q$ the only integration constant which describes the scalar hair. The boundary conditions, which are chosen as the asymptotic behavior of these solutions, preserve all the AdS symmetries. We want to point out that solutions described by (\ref{sl1}) only represent a line in the space of solutions. With these boundary conditions, we can integrate (\ref{Mw}) and get the mass defined by Wald's formula $M_W=0$.

Just like the previous soliton solutions, the two entropies are inconsistent. There is an integration constant $q$ which is not related to the mass, which means the massless state is degenerate (this conclusion is also made in~\cite{Henneaux:2002wm}) and should have a non-zero statistical entropy. On the other hand, these black hole solutions all have vanishing horizon, thus, have vanishing Bekenstein-Hawking entropy $S=0$.

\textit{To solve this entropy problem, we need another quantity to determine the macroscopic thermodynamic properties of these solutions, which we call a scalar charge}. Then we find that the mass depend on the scalar hair, in such a way that these solutions are not all indistinguishable massless black holes.

\section{Introducing a new scalar charge}\label{Sch}
\subsection{The existence of a new charge attached to scalar field}
The existence of the scalar charge has already been discussed not only in asymptotically AdS spacetimes ~\cite{Lu:2013ura,Lu:2014maa,Liu:2013gja}, but also in asymptotically flat spacetimes~\cite{Gibbons:1996af}. The authors of these papers assume the existence of such a charge because their definitions of mass are different from the Hamiltonian formula and Wald's formula, and the thermodynamic analysis need the scalar field to play the role as a charge.

For example, consider again the four-dimensional theory in section \ref{ss}. If we define the AMD mass~\cite{Ashtekar:1984zz,Ashtekar:1999jx} $M_A$ as the right mass (as in~\cite{Lu:2014maa}), then the variation of the non-integrable mass $M_W$ (or $M_H$) can be expressed as
\begin{equation}\label{Ni1}
\delta M_W=\delta M_A+\frac{4\pi}{3 \ell^2}(2 \phi_2 \delta \phi_1-\phi_1 \delta \phi_2)\,,
\end{equation}
with $M_A$ given by
\begin{align}
M_A=-8\pi \alpha\,.
\end{align}
This looks just like the non-integrable $\delta M_W$ (\ref{Mw2}) of the RN black holes. It is thus natural to consider the second term of the right hand side of Eq. (\ref{Ni1}) as a contribution from a scalar charge
\begin{equation}\label{Sc1}
\frac{4\pi}{3 \ell^2}(2 \phi_2 \delta \phi_1-\phi_1 \delta \phi_2)=-\Phi_S\delta Q_S\,,
\end{equation}
(we will give another definition of the scalar charge later). Wald's formula (\ref{Fl}), then gives the familiar first law
\begin{equation}
T\delta S=\delta M_A-\Phi_S\delta Q_S\,.
\end{equation}
This have the same formula as the first law of electrically charged RN black holes.

One interesting fact is that, as mentioned in~\cite{Lu:2013ura,Lu:2014maa,Anabalon:2014fla}, when the boundary conditions preserve all the AdS symmetries (see (\ref{boundarycondition})), the contribution from the scalar charge $\Phi_S\delta Q_S$ in (\ref{Sc1}) vanishes. We can see this directly from (\ref{Ni1}) and get $\delta M_W=\delta M_A$. It seems that these special boundary conditions makes $\delta M_W$ integrable and trustful, however in these cases the boundary conditions are also chosen before we calculate the mass, so $M_W$ still depends on the integration path.

The non-integrability of $\delta M_W$ not only happens in such Einstein-scalar gravities and RN-black holes, but also in many other gravities coupled to other matter fields~\cite{Lu:2013ura,Chow:2013gba,Liu:2014tra,Liu:2014dva,Fan:2014ixa,Fan:2014ala,Fan:2015yza}. \textit{These theories clearly show that when there exist charges which are not the Noether charge of any diffeomorphisms of spacetime along Killing vectors, the integral (\ref{Mw}) would be non-integrable}. Comparing with these theories, assuming the existence of a scalar charge is reasonable.

Another reason to assume the existence of the scalar charge $Q_S$ is that, the full space of solutions has two independent integration constants, so it is hard to believe that energy is the only charge. If the scalar charge $Q_S$ exists, the macroscopic thermodynamic properties of the solutions should be described by two parameters $M$ and $Q_S$, so the entropy $S(M,Q_S)$ would count the microscopic states which have mass $M$ and scalar charge $Q_S$. We will see this entropy can be consistent with the Bekenstein-Hawking entropy.

Before we give a new definition of mass, we need to define the scalar charge first. Since there is no symmetry corresponding to this charge, it cannot be defined as a Noether charge, thus, there may be some ambiguities to define the scalar charge. Here we define the scalar charge $Q_S$ as the coefficient of the leading term of the large $r$ expansion of the scalar field
\begin{align}
Q_S=\phi_1.
\end{align}

It is quite natural to define the scalar charge in this way. First, $\phi_1$ is an independent integration constant which describes the scalar hair. Second, when $\phi_1=0$ the scalar field vanishes everywhere, and also the contribution of the scalar field in $\delta M_W$ vanishes. At last, unlike the definition in (\ref{Sc1}), when we take $\phi_1$ as the scalar charge, its contribution in $\delta M_W$ has the standard form of a charge contribution $-\Phi_S\delta Q_S$, where we define the coefficient of $\delta Q_S$ multiplied by $-1$ as the conjugate potential $\Phi_S$.

As mentioned in~\cite{Gibbons:1996af}, this scalar charge is not conserved. We cannot construct a conserved current from a scalar field the way we construct the Bianchi identity from the gauge fields.
In~\cite{Gegenberg:2003jr,Park:2004yk,Anabalon:2015ija} the free energy of some exact scalar hairy black-holes has been calculated and compared with the free energy of the corresponding Schwarzschild AdS black holes at the same temperature. The results show that the Schwarzschild AdS black hole solutions are always thermodynamically preferred, which means the scalar hairy black holes will always decay into Schwarzchild AdS black holes at last. This is not surprising as we have mentioned that the scalar charge is not conserved.

\subsection{Solutions with a manifest scalar charge}
We now show a class of 3-dimensional scalar hairy soliton like solutions (the metric is regular everywhere, while the scalar field has a  $\log$ divergence at the origin) constructed in~\cite{Wen:2015xea}. These solutions have a manifest scalar charge term in the thermodynamic first law. The Lagrangian (\ref{Lag}) with a scalar potential
\begin{subequations}\label{V2}
\begin{align}
&V(\phi)=\frac{g^2\cosh^6 \psi}{64}\times
\nonumber
\\
&~~~~~~\big \{\cosh[6 \psi]-81\cosh[2 \psi]-6(7+\cosh[4 \psi])\big \},
\\
&\psi=\frac{\phi}{2\sqrt{6}},
\end{align}
\end{subequations}
admits the following solutions
\begin{subequations}\label{sl2}
\begin{align}
ds^2=&-\frac{g^2r^2(q^2+4qr+2r^2)}{2(q+r)^2}dt^2
\nonumber
\\
&+\frac{2 r^4 }{g^2(q+r)^4(q^2+4qr+2r^2)}dr^2+r^2d\theta^2\,,
\\
\phi(r)=&2\sqrt{6}~\text{Arctanh}\frac{1}{\sqrt{1+r/q}}.
\end{align}
\end{subequations}
Again, these solutions are not the whole space of solutions, and in fact on a integral path when we integrate $\delta M_W$. Applying Wald's formula we get the surface integral on the boundary
\begin{equation}
\delta M_W=\int_{S_{\infty}}(\delta Q_{\xi}-\xi\cdot\Theta)=\frac{2\pi}{ \ell^2}q \delta q,
\end{equation}
and also the surface integral on the infinitesimal ball around the origin
\begin{equation}\label{S0}
\int_{S_{0^+}}(\delta Q_{\xi}-\xi\cdot\Theta)=\frac{2\pi}{ \ell^2}q \delta q\,.
\end{equation}
The non-vanishing of the integral (\ref{S0}) usually would not happen in higher dimensions, but is possible in 3-dimensions as we can see. As the solutions have no event horizon, the integral (\ref{S0}) cannot be interpreted as $T\delta S$, so we interpret it as a contribution from scalar charge $\Phi_S\delta Q_S$ instead. This charge term is not an integral on the boundary, so even if we can accept a non-integrable mass, it cannot be absorbed by $\delta M_W$. Hence this scalar charge term is always manifest in the first law
\begin{equation}
\delta M_W=\Phi_S\delta Q_S.
\end{equation}

It may be inappropriate to call this equation a thermo-
dynamic first law, since there is no entropy, so the solution is not a thermodynamic
system. It describes how the other charges change with the scalar hair.

The existence of these solutions gives further evidence for the existence of the scalar charge.

\section{A new mass definition}\label{newmass}
In section \ref{Pro} we discussed the problems of the masses defined by the Hamiltonian formula and Wald's formula. They all define the mass on a variational level, and the results are non-integrable when matter fields are included. Choosing an integral path, $\delta M_W$ can be integrated and give a finite mass, however this mass depends on the integral path (or the boundary condition) we choose. The entropy problem of some solutions also indicates the masses defined in these formulas are not right. We think Wald's formula is not used properly by giving the definition of mass (\ref{Mw}). In this section, based on Wald's formula, we propose a new way to define the mass for theories including matter fields.

We first need to find out why the variation of mass calculated by (\ref{Mw}) becomes non-integrable when we include matter fields with charges, which are not the Noether charge of any diffeomorphism of spacetime along Killing vectors. Again we use the electrically charged RN-AdS black holes (\ref{RN}) as an example. \textit{Unlike the energy $M$ and angular momentum $J$, the electric charge $Q$ is not a Noether charge associated to a Killing vector in spacetime, instead, its existence and conservation is the result of an internal $U(1)$ gauge symmetry. When we calculate $\delta M_W$, we have chosen $\xi=\partial/\partial t$, thus other charges, for example the angular momentum $J$, which generate spacetime deffeomorphisms will not arise. This chosen of $\xi$ equals to stop the rotation of the solution,  however, it is not enough to stop the internal gauge transformations. As a result, the variation of electric charge $\delta Q$ will arise in the integration $\delta M_W$, thus makes it nonintegrable}. This can be seen clearly in (\ref{Mw2}).

The scalar charge we defined generates no transformations, so the arising of $\delta Q_S$ in $\delta M_W$ will also happen, as we can see from the previous sectors.

The right mass should be totally independent from not only charges like $J$ which generate spacetime diffeomorphisms, but also charges like $Q$ which are not attached to a spacetime Killing vector. Hence, when we calculate the variation of the mass, we should not only stop the other spacetime deffeomorphisms, but also the internal gauge transformations. In other words, charges like $Q$ and $Q_S$ should be considered as pure numbers without variation when we calculate the mass. 

Although the Hamiltonian formula and Wald's formula directly constructed the Hamiltonian, which seems to make their definitions of mass beyond controversy. However, they both define the mass on a variational level, which means we still have room to do some modification before we ruin physical fact that the mass is the Hamiltonian. Following the arguments above, the modification is to remove the contributions from  the variations of all other charges, which are not associate to a Killing vector, by hand. 

This can be done by defining a variation $\bar{\delta}$ which act only along the mass. By definition we have
\begin{equation}
\bar{\delta} Q=0\,,
\end{equation}
for the electrically charged RN-AdS black holes. The contribution from $\delta Q$ will be removed if we define the variation of the mass as $\bar{\delta}M_W$, which is given by
\begin{equation}
\bar{\delta} M_W=\bar{\delta} M-\Phi(M,Q)\bar{\delta} Q=\bar{\delta} M.
\end{equation}
So after we confine the $\delta$ in (\ref{Mw}) to be $\bar{\delta}$, Wald's formula can give the right mass $M$ for these RN black holes.

\textit{We now arrive at our new definition of mass $M_N$,
\begin{equation}\label{Nm}
\bar{\delta}M_N=\int_{S_\infty}(\bar{\delta} Q_{\xi}-\xi\cdot\Theta(\varphi,\bar{\delta}\varphi))\Big |_{\xi=\partial/\partial t}.
\end{equation}
with the variation $\bar{\delta}$ only along along $M_N$ in the space of solutions}. 

It is quite natural to define the mass in this way, because it uses the fact that the Hamiltonian should be a totally independent charge. After we integrate $\bar{\delta}M_N$, the result $M_N$ is in fact the real Hamiltonian.

To test this new definition, we can use it to calculate the mass of the scalar hairy black holes. For these black holes, as we do not agree with the mass defined by the Hamiltonian formula and Wlad's formula, we do not know what the real mass is before calculation. After we substitute into the solutions and perform some Legendre transformations, we can write the right hand side of (\ref{Nm}) as a total variation plus a charge term $-\Phi_S\bar{\delta}Q_S$. Then we apply $\bar{\delta}Q_S=0$, so only the total variation remains. This total variation is the variation of the real mass $\delta M_N$.

Based on the analysis in~\cite{Lu:2014maa}, we consider $d$-dimensional gravities minimally coupled to a scalar field with a scalar mass $m^2$ such that
\begin{equation}\label{md}
-\frac{1}{4\ell^2}(d-1)^2<m^2<-\frac{1}{4\ell^2}(d-1)^2+\frac{1}{4\ell^2},
\end{equation}
and a general scalar potential $V(\phi)$ admitting a Taylor expansion of the form
\begin{equation}
V(\phi)=-\frac{(d-1)(d-2)}{\ell^{2}}+\frac{1}{2}m^2\phi^2+\gamma_3\phi^3+\gamma_4\phi^4+\cdots.
\end{equation}
Let us apply (\ref{M1}) as the metric ansatz and define
\begin{equation}\label{sigma}
\sigma=\sqrt{4\ell^2m^2+(d-1)^2}.
\end{equation}
The asymptotics of the metric functions and scalar field will take the form
\begin{subequations}\label{sle}
\begin{align}
\phi &= \frac{\phi_1}{r^{(d-1-\sigma)/2}}+\frac{\phi_2}{r^{(d-1+\sigma)/2}}
 + \cdots \,,
\\
h&=g^2 r^2 + 1-\delta_{d,3} + \frac{\alpha}{r^{d-3}} +
\cdots\,,
\\
f &=g^2 r^2 + 1-\delta_{d,3} + \frac{b}{r^{d-3-\sigma}} + \frac{\beta}{r^{d-3}} +
\cdots\,,
\end{align}
\end{subequations}
with $\alpha$ and $\phi_1$ the only two independent integration constants. Substituting the expansions (\ref{sle}) into the equations of
motion and solving for the first few coefficients, we find that
\begin{subequations}
\begin{align}
&b=\frac{(d-1-\sigma)\, \phi_1^2}{4(d-2)\ell^2}\,,
\\
&\beta=\alpha + \frac{[(d-1)^2-\sigma^2]\, \phi_1\phi_2}{2(d-1)(d-2)\ell^2}\,.
\end{align}
\end{subequations}

Using Eq.(\ref{Wf}), the integration $\int_{S_\infty}(\bar{\delta} Q_{\xi}-\xi\cdot\Theta(\varphi,\bar{\delta}\varphi))$ on the boundary gives
\begin{widetext}
\begin{align}
\bar{\delta}M_N =& \omega_{d-2}\Big[
  -(d-2)\bar{\delta}\alpha +
\frac{\sigma}{2(d-1)\ell^2}\,
   [(d-1+\sigma)\phi_2\bar{\delta}\phi_1 - (d-1-\sigma)\phi_1\bar{\delta}\phi_2]\Big]
   \nonumber\\
=& \omega_{d-2}\Big[\bar{\delta}\left( -(d-2)\alpha - \frac{\sigma(d-1-\sigma)}{2(d-1)\ell^2}\phi_1\phi_2\right)+
\frac{\sigma}{\ell^2}\,
  \phi_2\bar{\delta}\phi_1\Big]
\nonumber\\
=&\bar{\delta}\left[\omega_{d-2}\left( -(d-2)\alpha - \frac{\sigma(d-1-\sigma)}{2(d-1)\ell^2}\phi_1\phi_2\right)\right]\,,
\label{deltaHgen}
\end{align}
\end{widetext}
where we apply $\bar{\delta}\phi_1=0$ in the third line of the equation.

The new integrable mass $M_N$ is given by
\begin{equation}\label{Mn1}
M_N=\omega_{d-2}\left( -(d-2)\alpha - \frac{\sigma(d-1-\sigma)}{2(d-1)\ell^2}\phi_1\phi_2\right).
\end{equation}

This new mass $M_N$ is different from the AMD mass which in these cases is given by
\begin{equation}
M_A=-\omega_{d-2}(d-2)\alpha.
\end{equation}

Our new mass is also different from the mass $M_C$ calculated by the ``counterterm subtraction method" . When there is a logarithmic $r$ dependence in the asymptotic expansions for the metric and scalar field, a logarithmic divergence will arise in the action. We need a boundary term from the scalar field
\begin{align}\label{boundaryL}
{\mathcal{L} }_{surf}[\phi] &= \frac{\gamma}{16\pi G} \sqrt{-h}\,
n^\mu\, \phi\,\partial_\mu\phi \,,
\end{align}
to remove this divergence and the parameter $\gamma$ can be fixed. Usually the result $M_C$ is also different from $M_A$.

However, under the condition (\ref{md}), no logarithmic function will arise in the asymptotic expansion for metric and scalar field thus there is no logarithmic divergence, and the contribution from (\ref{boundaryL}) also become finite. The parameter $\gamma$ now is free and the result $M_C$
\begin{align}
M_{C}=\frac{\omega_{d-2}}{16\pi}\Big[(2-d)\alpha +
  \frac{\left((d-1)(4\gamma-1)+\sigma\right)\, \sigma}{2(d-1)\ell^2}\, \phi_1\, \phi_2
     \Big]\,
\end{align}
depend on it, which means this method cannot give a unique mass. We can reproduce $M_A$ from $M_C$ by taking
\begin{align}
 \gamma=\frac{d-1-\sigma}{4(d-1)}.
\end{align}
And also we can reproduce our new mass $M_N$ by just taking
\begin{align}
\gamma=0,
\end{align}
which seems to be a more natural choice.

The details for the calculation of $M_C$ can be found in~\cite{Lu:2014maa}.

Although we modified the definition of mass, but the holding of Wald's formula (\ref{Fl}) is always a mathematical fact for on-shell spacetimes and would give the thermodynamic first law. Replacing $\bar{\delta}$ with the general variation $\delta$, the second line in (\ref{deltaHgen}) is just $\delta M_W$. Thus, Wlad's formula gives
\begin{align}
T\delta S=&\delta M_N+ \frac{\omega_{d-2}\sigma}{\ell^2}\phi_2\delta\phi_1
\nonumber\\
=&\delta M_N-\Phi_S\delta Q_S\,,
\end{align}
with the conjugate potential $\Phi_S$ given by
\begin{align}
\Phi_S=-\frac{\omega_{d-2}\sigma}{\ell^2}\,\phi_2\,.
\end{align}

The 3-dimensional massless black holes described by (\ref{V1}) and (\ref{sl1}) have a scalar mass $m^2=-\frac{1}{4\ell^2}(d-1)^2+\frac{1}{4\ell^2}=-\frac{3}{4\ell^2}$ which is outside (\ref{md}), but the solutions (\ref{sl1}) can still be described by (\ref{sle}) and do not contain any logarithmic functions, so the general analysis above is still valid here. The solutions (\ref{sl1}) have
\begin{align}
\phi_1=2\sqrt{2\mu q}\qquad \phi_2=-\frac{\sqrt{2\mu q^3}}{3}\qquad \alpha=0,
\end{align}
so according to (\ref{Mn1}), the masses for different $\mu$ are given by
\begin{equation}
M_N= - \frac{\pi}{2\ell^2}\phi_1\phi_2=\frac{2\pi\mu q^2}{3\ell^2}.
\end{equation}

Note that, $\mu$ is a parameter in the Lagrangian, so is not an integration constant and should have no variation. Using our definition of mass (\ref{Nm}), we see the solutions (\ref{sl1}) are actually massive black holes, and their masses are related to the scalar hair. As we have introduced a new macroscopic quantity for these black holes, the statistical entropy $S(M_N,Q_S)$ now counts the number of micro states with a mass $M_N$ and a scalar charge $Q_S$. The solution (\ref{sl1}) with
\begin{equation}
M_N=\frac{2\pi\mu q^2}{3\ell^2}  \qquad   Q_S=2\sqrt{2\mu q},
\end{equation}
is now unique, and could be considered as a quantum state, thus have a vanishing statistical entropy. Which is now consistent with its Bekenstein-Hawking entropy. This solves the entropy problem we presented in Section \ref{Pro}.

The newly defined mass $M_N$ is integrable and does not have the entropy problem mentioned above. The requirement fot the variation $\delta$ in (\ref{Mw}) to be $\bar{\delta}$ is based on the fact that the mass should be an independent charge of the solution. The consistence conditions (\ref{Re1}) and (\ref{Re2}) for the existence of a well defined energy is actually satisfied by $\bar{\delta} M_N$ rather than $\delta M_W$. Our definition (\ref{Nm}) is in fact consistent with the spirit of Wald's construction. The Hamiltonian formula can also be modified in this way as it also defines the mass on a variational level.

\section{Mass in more general gravities coupled to matter fields}\label{generalize}
Although the initial motivation of this paper is to give a well defined mass for asymptotically AdS Einstein-scalar gravities, our new definition of mass is valid for much more general gravities. The only requirements for the application of our definition can be conclude as: \textit{a), The solution should admit $\partial/\partial t$ as a Killing vector; b), All the terms in Wald's formula should be finite; c), All the independent charges can be identified.}

Consider, in general, a gravity coupled to matter fields $A_i$ with charges $Q_i$ which generate internal symmetry transformations and also matter fields $\Psi_j$ with charges $\psi_j$ which do not generate any symmetry transformation (for example the scalar charge we defined). Further more, if the Killing vector $\xi_k$ of solution have components along other directions of spacetime, there will be charges $J_k$ which generate the spacetime diffeomorphisms along those directions (for example, the angular momentum).

Wald's formula gives a relationship between the variations of all charges, and when the black hole can be considered as thermodynamic system, Wald's formula can be recognised as the thermodynamic first law. The $Q_i$ and $J_k$ charges are Noether charges and can be determined by the solution, while the $\psi_j$ charges are not. We need to give proper definitions for these $\psi_j$ charges the way we define the scalar charges. For a solution with all the charges mentioned above, and after some Legendre-like transformations, Wald's (\ref{Fl}) formula could be written in the following formula 
\begin{align}\label{Fl'}
T\delta S=\delta \mathcal{M}(\varphi)-\sum_{i}\Phi_i\delta Q_i-\sum_{i}\Phi_j\delta \psi_j-\sum_{k}\Phi_k\delta J_k
\end{align}
where the total variation $\delta\mathcal{M}$ is expected to depend on an extra integration constant which is independent from all the other charges (for example, the parameter $\alpha$ in (\ref{deltaHgen})), and may also depend on the other charges. Also we would find that $\delta M_W$ becomes non-integrable and can be written as
\begin{align}
\delta M_W=\delta \mathcal{M}(\varphi)-\sum_{i}\Phi_i\delta Q_i-\sum_{i}\Phi_j\delta \psi_j.
\end{align}
Terms proportional to $\delta J_k$ would not appear in $\delta M_W$ because we have chosen $\xi=\partial/\partial t$. 

It should be pointed out that, if we make another gauge choice that the gauge fields $A_i$ vanishes asymptotically on the boundary, rather than on the horizon, the contributions from $\delta Q_i$ in $\delta M_W$ will be moved to the surface integral on the horizon. For example, consider again the RN-AdS black holes, the surface integrals now becomes
\begin{align}
&\int_{S_{r_0}}(\delta Q_{\xi}-\xi\cdot\Theta)=T\delta S+\Phi(M,Q)\delta Q\,,
\\
\label{Mw2}
&\int_{S_{\infty}}(\delta Q_{\xi}-\xi\cdot\Theta)=\delta M=\delta M_W\,.
\end{align}
This changing of gauge choice may solve the non-integral mass problem of the RN black holes in some sence; however, this would not work for the scalar charges.

Following (\ref{Nm}) we get
\begin{align}
\bar{\delta} M_N=\bar{\delta} \mathcal{M}(\varphi)
\end{align}
where we have apply $\bar{\delta}Q_i=0$ and $\bar{\delta}\psi_j=0$.
This means $M_N=\mathcal{M}(\varphi)$.

These general analysis can be checked by looking at the results of some recent papers ~\cite{Lu:2013ura,Liu:2014tra,Liu:2014dva,Fan:2014ixa,Fan:2014ala,Fan:2015yza}. Using Wald's formula, these papers study the thermodynamic first law of gravities coupled to matter fields with both $Q_i$ and $\psi_j$ type of charges. And the first laws they get are all consistent with (\ref{Fl'}).

AdS dyonic black holes in gauged supergravities are constructed in~\cite{Lu:2013ura} with one pair of electric and magnetic charges, and in~\cite{Chow:2013gba} with more pairs of charges. The scalar fields in their solutions are totally determined by the electric and magnetic charges. This means the solutions in both~\cite{Lu:2013ura} and ~\cite{Chow:2013gba} are in fact not the full solutions. In our opinion, their scalar charge have been settled down and does not present as an independent integration constant. Facing a non-integrable Hamiltonian, authors of~\cite{Lu:2013ura} admit the existence of the scalar charge and take the AMD mass as the mass. While authors of~\cite{Chow:2013gba} think the mass is ill defined and does not exist, unless additional consistent boundary conditions are given by hand, which coincide with the mass defined on boundary conditions~\cite{Henneaux:2004zi,Hertog:2004dr,Henneaux:2006hk,Hertog:2005hm,Amsel:2006uf}.

We do not agree with both of them, because the requirements of the AMD method are not fully satisfied and the mass defined on the boundary conditions also have problems we presented in Sec. \ref{Pro}.

\section{Mass in a gravity with holography}
According to the AdS/CFT~\cite{Maldacena:1997re,Witten:1998qj,Gubser:1998bc} correspondence, the $(d-1)$-dimensional CFT generating functional of correlation functions for some operator $\mathcal{O}$ is equivalent to the partition function of a gravity in $AdS_d$ background with some specific boundary conditions. The Einstein-scalar gravities (\ref{Lag}) with a boundary condition $\phi_2(\phi_1)$ is dual to a CFT with some kind of multi-trace deformation~\cite{Berkooz:2002ug,Sever:2002fk,Witten:2001ua}.

Consider a CFT with a general multi-trace deformation $W(\hat{\mathcal{O}})$
\begin{align}\label{deformedcft}
S=S_{CFT}+\int \rho(x)~\hat{\mathcal{O}}+\int W(\hat{\mathcal{O}}),
\end{align}
with $\hat{\mathcal{O}}$ the dual operator of the scalar field and $W(\hat{\mathcal{O}})$ a general function of $\hat{\mathcal{O}}$. When the scalar mass satisfies $-\frac{1}{4}(d-1)^2<m^2\ell^2<-\frac{1}{4}(d-1)^2+1$, either the $\phi_1$ and $\phi_2$ mode are normalizable, thus, we can take either $\phi_1$ and $\phi_2$ as the source $\rho(x)$. If we take $\rho=\phi_2$, then the deformed CFT (\ref{deformedcft}) is dual to the Einstein-scalar gravity with the following boundary condition
\begin{align}
\phi_2(\phi_1)=\frac{\delta W(\phi_1)}{\delta \phi_1}.
\end{align}

As the boundary condition is determined by the boundary CFT (\ref{deformedcft}) and is added by hand in gravity, it seems a little unnatural to impose a boundary condition when someone only look at the gravity side. A more comforting way to interpret a boundary condition is to give additional special boundary terms (which is also determined by the boundary CFT) to the action of the gravity, thus the boundary condition is the solution of the boundary equations of motion~\cite{Sever:2002fk}. From this point of view, the same gravity with different boundary conditions are actually different gravities. Hence one would expect that a specific black hole would have different masses when its variation is confined by different boundary conditions, because the actions changes.

Following the spirit of this paper, we also propose a way to calculate the mass in a gravity with holography (or with additional boundary conditions). We should start from the new gravity action with the additional special boundary terms. The additional boundary terms may change the symplectic structure of the theory, thus give a different mass if we use our definition of mass proposed in Sec.~\ref{newmass}.

We cannot guarantee the validity of our definition of mass in this situation. The gravity with boundary conditions is still a subtle topic of gravity, since the boundary conditions cannot be naturally generated from only the gravity side.

\begin{acknowledgements}
I am grateful to Lin-Qing Chen, Si Li, Hai-Shan Liu for helpful discussions, to Andrzej Banburski, H. Lu, Rob Myers and Chris Pope for helpful comments and reading the paper, and especially to Wei Song for critical comments and numerous discussions on holography and boundary conditions. I also would like to thank Chuan-jie Zhu, the Beijing Municipal Education Committee and the Perimeter Institute Visiting Graduate Fellows program for financial support. Research at Perimeter Institute is supported by the Government of Canada and by the Province of Ontario through the Ministry of Research and Innovation. This work is also supported in part by NSFC 34112027.
\end{acknowledgements}


\end{document}